\documentclass{emulateapj}

\usepackage{array}

\begin{document}

\newcommand{\kms}{km~s$^{-1}$}
\renewcommand{\bf}{ }

\title{Hydrodynamics of high-redshift galaxy collisions:\\
From gas-rich disks to dispersion-dominated mergers and compact spheroids}

\author{Fr\'ed\'eric Bournaud, Damien Chapon, Romain Teyssier, Leila C. Powell}
\affil{Laboratoire AIM Paris-Saclay, CEA/IRFU/SAp -- CNRS -- Universit\'e Paris Diderot, 91191 Gif-sur-Yvette Cedex, France. frederic.bournaud@cea.fr}

\author{Bruce G. Elmegreen}
\affil{IBM T. J. Watson Research Center, 1101 Kitchawan Road, Yorktown Heights, New York 10598 USA, bge@us.ibm.com}

\author{Debra Meloy Elmegreen}
\affil{Vassar College, Dept. of Physics \& Astronomy, Poughkeepsie, NY 12604, elmegreen@vassar.edu}

\author{Pierre-Alain Duc}
\affil{Laboratoire AIM Paris-Saclay, CEA/IRFU/SAp -- CNRS -- Universit\'e Paris Diderot, 91191 Gif-sur-Yvette Cedex, France.}


\author{Thierry Contini, Benoit Epinat}
\affil{Laboratoire d'Astrophysique de Toulouse-Tarbes, Universit\'e de Toulouse, CNRS, 14 Avenue Edouard Belin, 31400 Toulouse, France.}

\author{Kristen L. Shapiro}
\affil{Department of Astronomy, Campbell Hall, University of California, Berkeley, CA 94720.\\
Aerospace Research Laboratories, Northrop Grumman Aerospace Systems, Redondo Beach, CA 90278, USA.}

\begin{abstract}
Disk galaxies at high redshift ($z \sim 2$) are characterized by high
fractions of cold gas, strong turbulence, and giant star-forming
clumps. {\bf Major mergers of disk galaxies at high redshift should then generally involve such turbulent clumpy disks. Merger simulations, however, model the ISM as a stable,} homogeneous, and thermally pressurized medium. We present the first merger simulations with high fractions of cold, turbulent, and clumpy gas. We discuss the major new features of these models compared to models where the gas is artificially stabilized and warmed. Gas turbulence, which is already strong in high-redshift disks, is further enhanced in mergers. Some phases are dispersion-dominated, with most of the gas kinetic energy in the form of velocity dispersion and very chaotic velocity fields, unlike merger models using a thermally stabilized gas. These mergers can reach very high star formation rates, and have multi-component gas spectra consistent with SubMillimeter Galaxies. Major mergers with high fractions of cold turbulent gas are also characterized by highly dissipative gas collapse to the center of mass, with the stellar component following in a global contraction. The final galaxies are early-type with relatively small radii and high Sersic indices, like high-redshift compact spheroids. The mass fraction in a disk component that survives or re-forms after a merger is severely reduced compared to models with stabilized gas, and the formation of a massive disk component would require significant accretion of external baryons afterwards. {\bf Mergers thus appear to destroy extended disks even when the gas fraction is high, and this lends further support to smooth infall as the main formation mechanism for massive disk galaxies.} 
\end{abstract}

\section{Introduction}

In the $\Lambda$-CDM cosmological model, collisions and mergers are an important growth mechanism for galaxies, and most of them should have occurred at high redshift ($z > 1$, e.g., \citealt{stewart09}). {\bf Major mergers, where the ratio of the baryonic masses of the involved galaxies is close to one, are in particular important for the build-up of massive galaxies \citep{hopkins10}.} Even if recent models and high-redshift observations suggest that the growth of galaxies near $L*$ is largely through relatively smooth and cold accretion \citep{dekel09, BE09, conroywechsler09, genel10}, about one-third of a galaxy's baryons are still expected to be provided by significant mergers \citep[e.g.,][]{DSC09,brooks}.

The effect of the merger of two disks galaxies has been largely
explored in the low-redshift context, where galactic disks are mostly
made up of stars with relatively modest fractions of cold gas. In such
conditions, it is known that a major merger of two spiral galaxies
typically produces a remnant resembling an elliptical or early-type
galaxy (ETG) \citep{BH96, MH96, naab03, bournaud04, naab06, naab07, bournaud07, hoffman10}. There are still some differences between the predictions of merger models and observations \citep{burkert-sauron} but there is, overall, an ample consensus that hierarchical merging is the main formation channel for present-day ETGs.

At high redshift, the role of major mergers in the formation of particular galaxies and in the growth of galaxies in general remains much more debated. First, determinations of the merger rate from high-redshift surveys are usually based on irregular structures or disturbed kinematics \citep[e.g.,][]{conselice,lotz06, lotz08, overzier, jogee09,
lopez-sanjuan09,puech}. However, primordial galaxy disks can have irregular morphologies and disturbed kinematics even when they are {\bf mostly isolated, not undergoing strong interactions} \citep{E07, E09,genzel08,BEE07,genzel08,bournaud08,DSC09}, so the actual merger rate remains unknown. Also, high-redshift disk galaxies have typical morphologies that differ from nearby spirals \citep{cowie96, EE05, E07, genzel08}, and their mergers could also have different morphologies and kinematics compared to nearby mergers, potentially making the identification of mergers more ambiguous: for instance, asymetries and clumpiness in optical imaging are often attributed to mergers \citep{conselice, lotz08}, but high-redshift disk galaxies seem to frequently form massive clumps just by internal  instabilities that do not require interactions \citep{E07,E09, genzel08}.
Even in cosmological simulations, the galaxy merger rate
is continuously revised \citep[e.g.,][and references
therein]{genel08,genel10}. {\bf In particular, mergers of dark
halos do not necessarily lead to the merging of their central galaxies,
and the baryonic mass ratio of galaxy mergers can significantly differ from the dark matter mass ratio of their host haloes \citep{hopkins10}.}

Second, the outcome of major mergers is usually considered to be
early-type galaxies (ETGs). However, high-redshift ETGs are often
unexpectedly compact compared to nearby ellipticals and merger
simulations \citep{daddi05,trujillo06,vdk, vandokkum08,kriek}. The role
of mergers in forming these ETGs has been questioned: while their
evolution into modern, more extended elliptical galaxies can be
explained through continuous hierarchical merging \citep{naab09}, the
emergence of compact ellipticals at high redshift remains more
mysterious.

Third, the most general issue is the ability of disk-dominated galaxies
to survive violent mergers. While it is known that mergers of disk
galaxies {\em can } produce ETGs with no or little residual disk
component (references above), whether all mergers even with very high gas fractions destroy disk galaxies and transform them into ETGs, or whether disk-dominated galaxies could survive major mergers in high-redshift conditions, remain more uncertain. The frequency of violent mergers could indeed call into question the survival of disk-dominated galaxies \citep[][]{weinzirl}. Some models of mergers, however, have led to the proposal that high-redshift merger remnants could be dominated by large rotating disk components rather than bulges and spheroids \citep{springel, robertson-mergers}. There have been claims that such disky merger remnants cannot account for the observed properties of high-redshift disk galaxies \citep{shapiro08,BE09} but this remains actively debated in simulations \citep{robertson-bullock} and in observations \citep{hammer}.

Thus, at least three major unknowns remain: the frequency and
observational signatures of high-redshift mergers, their role in
forming the high-redshift populations of ETGs, and the ability of
massive disk-dominated galaxies to survive these events.

\smallskip

The dynamics and outcome of mergers have never been
modeled for realistic, high-redshift galaxies. Simulations by
\citet{cox}, \citet{springel}, and \citet{robertson-mergers} considered
high gas fractions and concluded that disks could survive or re-form
after a merger. However, these simulations had high thermal pressure
support in the ISM (temperature$>$$10^4$K), which forces the gas to be smooth and stable in the isolated galaxies and in the colliding pairs
as well.

Star-forming galaxies at high redshift ($z =1 - 5 $) are very different
from such models. They are generally gassy, clumpy, and turbulent,
with $\sim50$\% of their baryons in cold molecular gas \citep{daddi10, tacconi10}, giant star-forming clumps of $10^7-10^9\;M_\odot$
\citep{E04,E07,E09}, and velocity dispersions of several tens of
km~s$^{-1}$ \citep{FS09, shapiro08, genzel08, epinat09, queyrel09}.
Associated star formation rates (SFRs) are around 100~M$_{\sun}$~yr$^{-1}$. {\bf While most observations of high-redshift star forming galaxies concerned only very massive galaxies (above the critical mass $L*$), recent studies of $z>1$ lensed galaxies find that gravitational instabilities and turbulent clumps could also be dominant in lower-mass galaxies \citep{jones, swinbank}. }
The morphology and dynamics of the disks suggest that the observed clumps form by gravitational instabilities \citep{E04, EE05,shapiro08,bournaud08, daddi10}. Simulations of this process
require cosmological codes that allow for highly supersonic turbulent speeds \citep{agertz09,ceverino09}. Even low-redshift spiral galaxies have an unstable, cloudy ISM supported mostly by supersonic turbulence rather than thermal pressure
\citep[e.g.,][]{burkert-ism}.  Recent simulations showed that a spatial resolution
of 10\% of the scale height with full tracking of turbulent motions
is required to reproduce the three-dimensional inertial range of
turbulence inside a modern galaxy \citep{bournaud10}.

A typical wet merger at high redshift should involve the interaction of such clumpy, turbulent disks. The outcome could differ from previously modeled mergers using thermally-stabilized gas.  Recent studies have shown that ISM turbulence and clumpiness have a substantial effect in mergers of present-day spirals with just a few percent of gas, with significant differences both in the star-formation history during the interaction \citep{teyssier10,saitoh} and the properties of the relaxed post-merger ETG \citep{BDE08, bois10}. The effect could presumably be more dramatic in high-redshift mergers involving high fractions of cold gas.

\smallskip

In this paper, we present hydrodynamical AMR simulations of major
mergers between galaxies with high fractions of cold, turbulent and
clumpy gas, like typical $z \sim 2$ star-forming galaxies. Gas cooling
below $10^4$~K is allowed, supersonic ISM turbulence is captured, and
the main star-forming complexes are directly resolved. We compare with
models using artificially stabilized disks, as in traditional
high-redshift studies.

We find that mergers of high-redshift disks can undergo a very clumpy
and chaotic phase, during which the kinematics is dispersion-dominated
even for the gas component. The gas kinematics appears consistent with
the observed properties of some ``dispersion-dominated galaxies'' as
well as the spectral properties of SubMillimeter Galaxies (SMGs). A large part of the total mass
is thus supported by supersonic turbulence, which is a dissipative
support, hence the merging system undergoes a rapid dissipational
collapse into a compact ETG. The mass fraction in a surviving or
re-formed disk component is much lower for supersonically turbulent high-redshift mergers than for models with artificially warmed gas. This does not mean that mergers of gas-rich galaxies can never re-form a disk: a large enough far-outer reservoir of gas could recollect into a new disk, or continuous infall of fresh gas could build a new disk. Some orbital parameters not modeled here might also be more favorable to disk survival \citep[e.g.][]{hopkins09}. Still, our main conclusion is that mergers between gas-rich turbulent and clumpy galaxies are qualitatively different than mergers between gas-rich smooth galaxies with artificial pressure support, and can better explain resolved observations of high-redshift systems.

\section{Simulations}
{\bf This paper studies idealized models of disk galaxies. The initial conditions are not cosmological, and there is no cosmologically-motivated boundary condition modeling mass infall or extended gas reservoirs. Instead, the initial conditions are designed to be representative of gas-rich major mergers of redshift $z \sim 2$ galaxies, based on their main observed characteristics. We model collisions and mergers between two such galaxies and study the effect of this merger, by comparison to the modeled evolution of a single isolated galaxy.}

The simulations were performed with the AMR code RAMSES
\citep{teyssier02}. The technique is fully described elsewhere and
shown to model realistic interstellar gas in low-redshift mergers and
isolated disks \citep{teyssier10,bournaud10}. The main parameters
adopted for the present simulations are in Table~1.

\begin{table*}
\centering \caption{Parameter used for the simulations.}
\begin{tabular}{lccl}
\hline
\hline

Model & Orbit & EoS & Comment   \\
\hline
C1 & 1  & cooling & merger of gas-rich clumpy disks        \\
C1F & 1  & cooling & C1 with stronger feedback      \\
C1F5 & 1  & cooling & C1 with 500\% feedback      \\

S1 & 1  & stabilized & gas-rich, smooth, stabilized disks     \\
S1F & 1  & stabilized & S1 with stronger feedback       \\
S1F5 & 1  & stabilized & S1 with 500\% feedback      \smallskip \\

C2 & 2  & cooling & merger of gas-rich clumpy disks     \\
S2 & 2  & stabilized & gas-rich, smooth, stabilized disks    \smallskip \\

C3 & 3  & cooling & merger of gas-rich clumpy disks       \\
S3 & 3  & stabilized & gas-rich, smooth, stabilized disks    \smallskip \\

I & isolated  & cooling &control run: high-redshift disk     \smallskip \\

LM-C2 & 2  & cooling & lower-mass gas-rich clumpy disks    \\
LM-C2F & 2  & cooling & C2 with stronger feedback    \\
LM-S2 & 2  & stabilized & lower-mass, smooth stabilized disks    \\

\end{tabular}
\end{table*}

\begin{table*}
\centering \caption{Properties
of the final relaxed systems:} Half-mass radius $R_{\mathrm{1/2}}$ and
Sersic index $n$ (best fit between 0.3 and 3 $R_{\mathrm{1/2}}$); gas fraction remaining at the pericenter and 400~Myr later; mass-weigthed average of the 1D local gas turbulent speed  $< \sigma_{\mathrm{turb,1D}}>$ measured 140~Myr after each pericenter;  baryonic mass fraction of the baryons in a rotating disk $f_{\mathrm{disk}}$, measured using a kinematic disk+spheroid decomposition as in \citet{MB10}.
\begin{tabular}{lcccccccccc}
\hline
\hline
 &  & \multicolumn{2}{l}{Gas consumption} && ISM turbulence && \multicolumn{3}{l}{Merger remnant properties} \\

Model  & \hspace{.2cm}  &  $f_{\mathrm{gas,peri}}$ & $f_{\mathrm{gas,i+400Myr}}$ & \hspace{.2cm} & $< \sigma_{\mathrm{turb,1D}}>$ (km\,s$^{-1}$)  & \hspace{.2cm}  &Sersic $n$ & $R_{\mathrm{1/2}}$ & $f_{\mathrm{disk}}$ \\
\hline
C1    &&  45\% & 17\%  && 175  &&4.4 & 2.8 &  12\%     \\
C1F   &&   54\%  & 22\% &&    186   &&4.2 & 3.0 &  16\%       \\
C1F5   &&   58\%  & 21\% &&    204   &&3.9 & 3.1 &  21\%       \\
S1     && 48\% & 20\% &&  56  &&2.9 & 6.1 & 40\%      \\
S1F     && 52\%   & 22\% &&  59  &&2.5 & 6.3 & 48\%    \\
S1F5     && 56\% & 23\% &&  65  &&2.6 & 6.2 & 53\%      \smallskip \\

C2     && 53\%  & 13\% &&     153   &&4.2 & 2.9& 14\% \\
S2     && 49\% & 18\% &&  68  && 3.2 & 5.8 & 37\% \smallskip \\

C3    &&  53\% & 19\%  &&     164 && 4.9 &  2.6  &  11\%    \\
S3    &&  47\% & 17\%  &&  72  &&  3.3&  5.6   &  37\%   \smallskip \\

I    && -- & --  &&       59  && 1.7 & 5.4 & 66\%    \smallskip \\

LM-C2      && 52\% & 19\% &&  89  && 3.7 & 1.2 & 10\% \\
LM-C2F      &&  58\% & 21\% &&  96  && 3.9 & 1.3 & 13\% \\
LM-S2     && 57\% & 22\% &&  42   &&2.7 & 2.0 & 35\% \smallskip \\

\end{tabular}
\end{table*}

\begin{figure*}
\centering
\includegraphics[width=15cm]{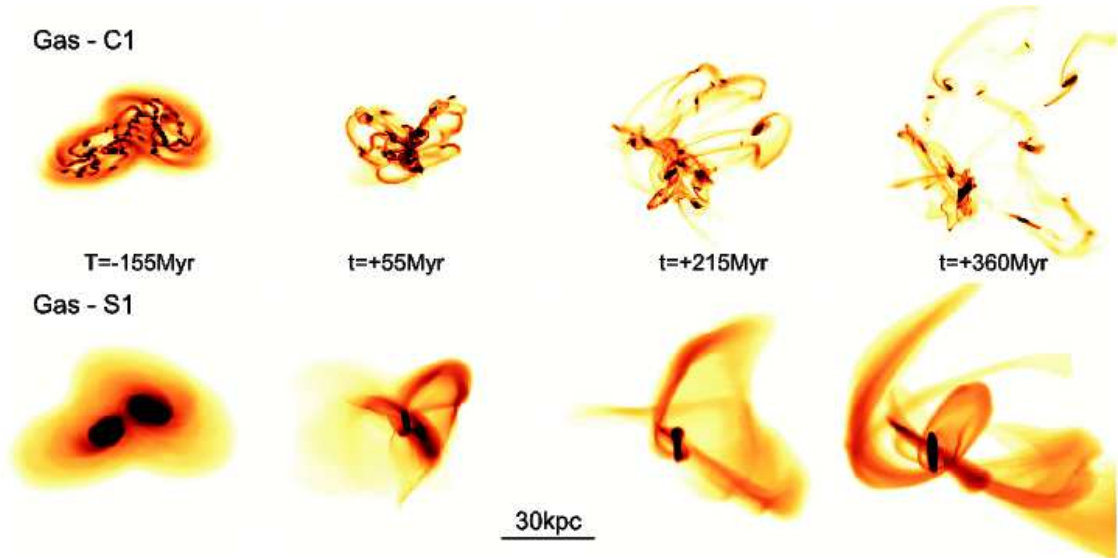}
\caption{Snapshots of the gas mass distribution for models C1 (cooling)
and S1 (stabilized ISM) at similar instants and under similar projections.
$t=0$ is the first pericenter passage and all maps are in log scale.}
\label{fig1a}
\end{figure*}

\begin{figure*}
\centering
\includegraphics[width=15cm]{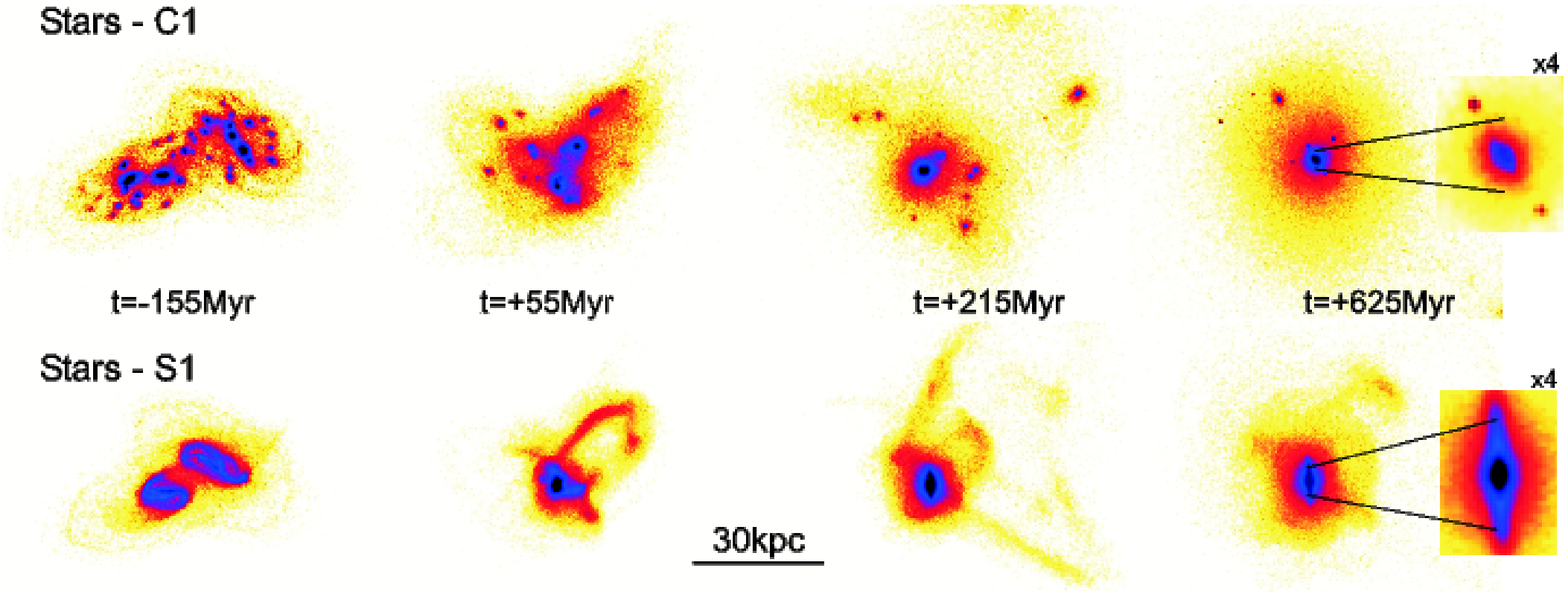}
\caption{Snapshots of the stellar mass surface density distribution for
models C1 (cooling) and S1 (stabilized ISM) at similar instants and under
similar projections. $t=0$ is the first pericenter passage and all maps
are in log scale.} \label{fig1b}
\end{figure*}

The box size for the simulation is 200~kpc. The coarsest level of the
AMR grid is $l=8$, which corresponds to a $256^3$ Cartesian grid with a
cell size of 781~pc. A cell that contains a gas mass larger than
$m_{\mathrm{res}} = 8 \times 10^4$~M$_{\sun}$, or a number of particles
larger than 15, is refined, until the maximal level $l=11$ is reached.
At that point, the cell size corresponds to 97~pc. Stars and dark
matter are described with $2 \times 10^5$ collisionless particles each
in the initial galaxies.

Our clumpy disk simulations were performed with a barotropic cooling
model \citep{bournaud10}, naturally producing a cloudy and
turbulent ISM, with a temperature floor around $10^3$~K. We refer to
these as the {\em cooling } models. For comparison, the same gas-rich
mergers were modeled with a thermally pressurized and Toomre-stable
ISM, using an adiabatic Equation of State (EoS) with an exponent of
$\gamma = 5/3$ for densities above 1~cm$^{-3}$, and $T$=$10^4$~K for
lower densities. This EoS maintains a Toomre parameter $Q\simeq 1.5-2$
in the initial disks and stabilizes the gas against axisymmetric
perturbations.  These will be called the {\em stabilized } models. In
all cases, a density-dependent pressure floor ensures that the Jeans
length is resolved by at least 4 cells to avoid artificial
fragmentation, as initially proposed by \citet{machacek} (see
\citealt{teyssier10} for the implementation details).

In the cooling models, star formation is assumed to proceed above a
density threshold of 300~cm$^{-3}$ with an efficiency of 7\%, i.e. 7\%
of the gas in a given cell forms stars per local free-fall time. This
gives a star formation rate (SFR) of 120~M$_{\sun}$~yr$^{-1}$ in the
isolated disk model, realistic for such gas-rich $z \sim 2$ systems. In
the stabilized models, the star formation threshold is 3~cm$^{-3}$ and
the efficiency is 4\%, giving the same SFR in the isolated disks: we
compare the two sets of models with the same SFR in pre-merger disks.
The resulting star formation history in the merger models with cooling/stabilized EoS is relatively similar (see Section~\ref{sec:sfr}).

All simulations use the kinetic feedback model described in
\citet{dubois}, with 20\% of the energy from each supernovae
re-injected in the form of an expanding bubble of initial radius
100~pc {\bf (assuming that the rest of the supernovae energy was radiated away before propagating up to scales of 100~pc)}. Some models have increased feedback with 100\% of the supernova energy injected into the ISM (models labeled "F") {\bf and some have a supernova feedback efficiency set to 500\% (models labeled "F5", discussed only in Section~3.4)}.

Our simulations start with disk galaxies, each having a stellar mass of
$4 \times 10^{10}$~M$_{\sun}$, an initial gas fraction of 70\%, a disk
scale length of 5~kpc with a truncation radius of 12~kpc, and an
initial scale height of 800~pc. An initial bulge containing 15\% of the
stellar mass is assumed, with a Hernquist profile and a scale length of
500~pc. The dark matter halo has a Burkert profile with a core radius
of 8~kpc, and a mass fraction (dark/total) inside the disk radius of
30\%. The circular velocity in the outer disk is 245~\kms.

In the cooling models, the initial disks spontaneously become clumpy
and turbulent, with $V/\sigma \simeq 5$ (for rotation speed $V$ and
turbulent speed $\sigma$). When the mergers occur, at the first
pericenter, gas fractions are around 50\% after some early gas
consumption (see Table~2): this seems typical for disk galaxies at $z
\sim 2$ \citep{daddi10, tacconi10}, and hence expected for a wet merger
at $z \sim 2$.

Our merger models start with relatively massive disk galaxies, so as to
compare with existing models, e.g. in \citet{robertson-mergers}. There
are galaxies like this in $z \sim 2$ samples (for instance in the
sample of \citealt{FS09}).  Also, our
pre-merger disks are not extreme in terms of gas clumpiness and
turbulence: many $z\sim 2$ disks have even lower $V/\sigma$ ratios
\citep[e.g.,][]{FS09}.

\medskip

Three interaction orbits were used. They are all prograde for one
galaxy and retrograde for the other in order to consider a typical
total angular momentum rather than extreme cases with aligned
spins\footnote{Aligned spins would not necessarily result in higher angular
momentum or more disk component in the final result, because of tidal
removal of the angular momentum \citep{robertson-mergers}}. Orbit~1 has
an impact parameter of 12~kpc and is parabolic. One disk has an initial
inclination of $30^\circ$ from the orbital plane, and the other has an
initial inclination of $50^\circ$. For orbit~2, these parameters are:
35~kpc, hyperbolic (total energy of the galaxy pair is 0.3 times its
initial kinetic energy), and $30^\circ$ and $50^\circ$ degrees,
respectively. For orbit~3, they are 25~kpc, hyperbolic (total energy
$=0.2\times$ the initial kinetic energy), $40^\circ$, and $65^\circ$
degrees, respectively.

Three simulations were also run for a merger of lower-mass galaxies. The masses of all initial components (stellar disk, gaseous disk, bulge, halo) where reduced by a factor of 8 compared to the initial higher-mass models, and all sizes and distances were reduced by a factor of $\sqrt 8$, hence keeping surface densities about constant. The pre-merger galaxies then had circular velocities of about 150~km~s$^{-1}$ in their outer disks, with a $V/\sigma \simeq 3.5$ ratio -- somewhat lower than in the more massive disks models, which is in agreement with observed trends \citep{FS09}. These merger models were performed using Orbit~2 with our cooling and stabilized ISM models, and are respectively denoted LM-C2 and LM-S2, and LM-C2F with increased feedback. 

\section{Results}
\begin{figure}
\centering
\includegraphics[width=8.5cm]{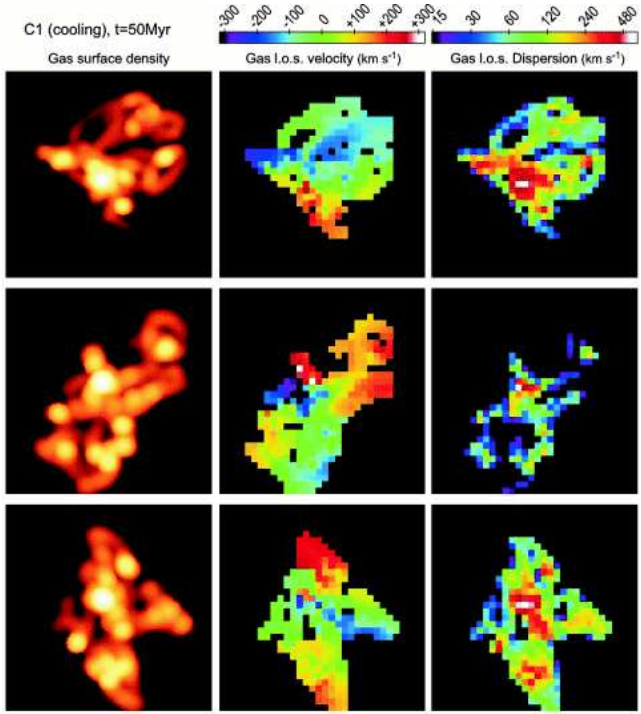}
\caption{Three orthogonal projections of model C1 showing the surface
density, velocity and dispersion maps (mass-weighted values along the
line-of-sight, for gas denser than 10~cm$^{-3}$). A resolution of 1~kpc
was assumed (FWHM of a gaussian beam). All snapshots are 44 $\times$
44~kpc. Note that H$\alpha$ observations would be mostly sensitive to
emission from the dense clumps.} \label{fig2}
\end{figure}

\begin{figure}
\centering
\includegraphics[width=8.5cm]{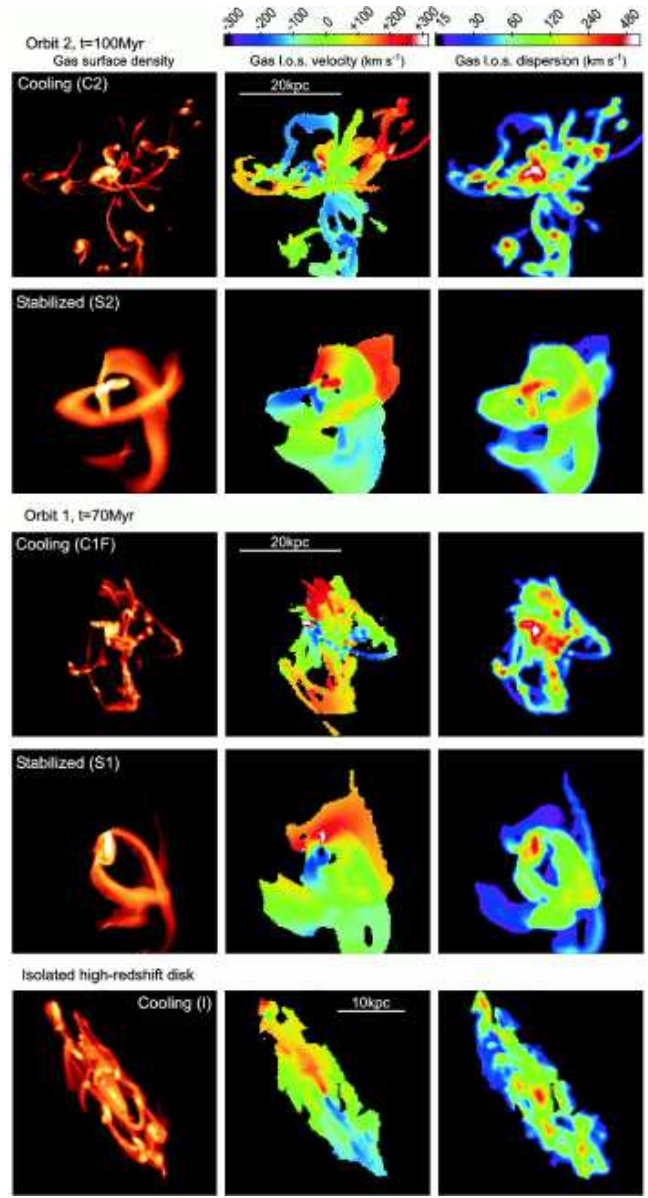}
\caption{Same as Fig.~2, with comparison of on-going mergers in the
cooling and stabilized ISM models at the same instant and under the same
projection, shown here at full resolution. An isolated clumpy turbulent
disk is also shown after a similar evolutionary time as the mergers. }
\label{fig3}
\end{figure}

\begin{figure*}
\centering
\includegraphics[width=12cm]{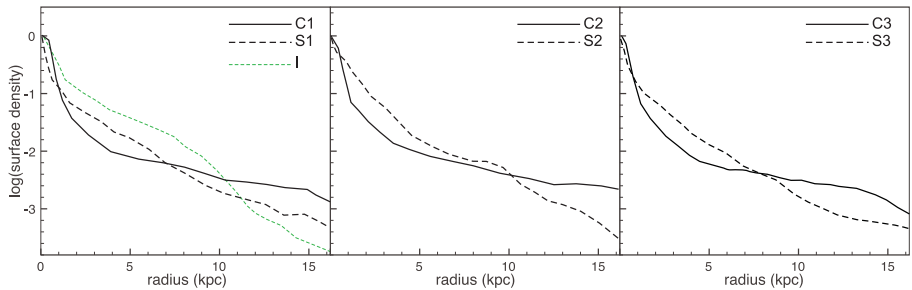}
\caption{Radial profiles of the stellar surface densities in final
merged and relaxed systems, and the isolated disk model after the same
evolutionary time. Several projections were averaged for each model.
Ellipse-fitting was performed with the {\tt ellipse} task in {\tt
IRAF}. } \label{fig4}
\end{figure*}

\subsection{On-going mergers}

The time evolution of mergers C1 (cooling) and S1 (stabilized ISM) is shown
in Figures~\ref{fig1a} and \ref{fig1b}. The stabilized model has
relatively homogeneous tidal tails orbiting around smooth gas disks.
The gas in the cooling model is dominated
by numerous star-forming clumps that are somewhat more massive than the
clumps were in the pre-merger galaxies. Visual inspection shows that
some of the pre-existing clumps survived and some have disrupted and
re-formed. Numerous filaments form in many directions, instead of a
main pair of continuous tidal tails. Other projections
(Fig.~\ref{fig2}) show that the gas distribution during the merger
forms a very irregular spheroid in three dimensions; there are no
disks. The model with stronger feedback (C1F in Fig.~3) is
qualitatively similar: self-gravity is already sufficient to drive strong
ISM turbulence in massive galaxies at high redshift \citep{DSC09,EB10}.

Gas velocity fields and line-of-sight dispersion maps are shown in
Figures~\ref{fig2} and \ref{fig3}. The gas velocity fields of our
high-redshift cooling models are very chaotic, and often lack extended
rotating components, such as large disks or long tidal tails with
monotonic velocity gradients. The gas velocity dispersions are high,
especially near dense clumps.

Interactions at low redshift substantially increase the gas velocity
dispersion, from typically 10~km~s$^{-1}$ in non-interacting spirals to
30--40~km~s$^{-1}$ in major mergers (see \citealt{E95} for
observations, \citealt{BDE08} for simulations). We find a similar
relative increase here, but starting with disks that are already quite
turbulent before the merger. This results in systems where the gas
component is dispersion-dominated during the merger, with $V/\sigma$
ratios\footnote{Throughout the paper, $\sigma$ refers to a
one-dimensional dispersion. A system with $V/\sigma < 2$ is
dispersion-dominated with $>60$\% of its kinetic energy support is in
the form of dispersions.} of around 2 or even 1 for some projections
and times. The gas dispersions are high in projection on the
line-of-sight, and also in direct measurements of the local
three-dimensional motions: we measured the turbulent velocity
dispersion in 1~kpc$^3$ cubes in model C1 at $t=140$~Myr; the
mass-weighted average value was $< \sigma_{\mathrm{turb, 1D}}> \simeq 175$~km~s$^{-1}$ (higher values on projected maps such as Figures~3 and 4 result from line-of-sight projection effects, not to a physical, local turbulent motion). Similar measurements are given for all models in Table~2, at the same instant which is about the peak of turbulent speed in our merger models with cooling. These measures show that the turbulent velocities become much higher during the interactions and mergers than in the pre-merger clumpy disks, and that this property is not reproduced in the simulations using a stabilized gas model. {\bf The values of line-of-sight velocity dispersions in Figures~3 and 4 will be further compared to high-redshift observations in Section~4.1.}

Mergers with the stabilized ISM model show velocity dispersions that
are lower by a factor $\sim 3$ for all times. Instead of chaotic
velocity fields, they have relatively smooth and extended rotating gas
disks, surrounded by long tidal tails that co-rotate with large-scale
velocity gradients (see examples in Fig.~3).

Low-redshift mergers have different signatures, since they typically
exhibit long tidal tails with large and smooth velocity gradients along
these tails (see examples in \citealt{bournaud-fp, chilingarian}).
Low-redshift merger models with warm stable gas versus colder and
unstable gas in cooling models have been compared in
\citet{teyssier10}. While they display substantial differences in their
star formation properties, both models are dominated by long tidal
tails and have similar half-mass sizes after the merger. The tails have star-forming clumps in the cooling model, but the clumps do not dominate the morphology and kinematics as they do at higher redshift.

In this sense, stabilized ISM models are a more acceptable recipe for the large-scale gas properties in low-redshift mergers than in high-redshift ones. This can be explained because low-redshift systems have weaker gas turbulence and are more stable than high-redshift systems on scales of a kpc. Low-redshift disks are only slightly unstable, and the sizes and masses of the clumps that form are much smaller than they are at high redshift. Thus the effects of ISM turbulence and clumpiness are more striking in high-redshift mergers. Resolving gas turbulence, star forming instabilities and/or shocks change the star formation history of these low-redshift mergers \citep{barnes04,saitoh, teyssier10,chien10}, but have no major impact on the morphology of the merger remnants and only a limited impact on their kinematics \citep{bois10}.

\subsection{Merger remnants}

Supersonic gas turbulence dissipates rapidly, which can produce a
strong contraction of a dispersion-supported gaseous system. Such
dissipative collapse is observed in Figure~1 for model C1. Some clumps
are expelled by tidal interactions, but most of the gas clumps and
inter-clump gas coalesce in a compact central object. Because the gas
fraction is initially high, a large part of the final stellar content
forms within the same gas clumps and follows their coalescence. 
Old stellar components contract due to the dissipation of the gas and young stars, which contain a large fraction of the total mass. Thus, the merger leads to the formation of a relatively compact stellar spheroid. This spheroid has a low half-mass radius, a high Sersic index, and an extended stellar halo {\bf visible as a low surface mass density component beyond radii of $\sim 7$~kpc in Figure~5}. Table~2 gives half-mass radii and Sersic indices for all models. Figure~4 shows the corresponding stellar mass profiles. Remnants of the
stabilized-ISM merger models are more extended\footnote{\bf Merger remnants in the stabilized models are more extended in terms of half mass radii: the outer, low-mass density halo is more extended in the cooling models, but would be hardly detected in the majority of high-redshift observations, as shown by Mancini et al. (2010).} and have lower Sersic indices than remnants of the cooling merger models. The isolated clumpy turbulent disk model has formed a central bulge through internal evolution (as in \citealt{noguchi}, \citealt{immeli} and \citealt{EBE08} models), but remains dominated by a rotating exponential disk (\citealt{BEE07}).

The merger remnant of a typical high-redshift merger is not just more
compact, but also less disky. Gas disk re-growth is faster in the
models with a stabilized ISM (see Fig~1 at $t=215$), and thus forms a
more massive final stellar disk. For example, the final disk component
in model S1 contains 40\% of the baryons and is obvious up to large
radius in Figure~1. In cooling model C1, most star formation takes
place in gas clumps before they coalesce and get a chance to re-form a
disk; the final disk in model C1 is compact, with a disk fraction of
only 12\%. Similar differences are found for all merger orbits (Table~
2).

\medskip

The increased dissipation and resulting compactness in cooling models with a
clumpy turbulent gas can be explained:  clumps interact with each other
gravitationally, scatter, and develop 3D motions (i.e., high
dispersions). The gas clumps and holes can then pass by each other.
In stabilized ISM models, the smooth distribution of the gas does not allow it to pass through itself and it therefore rapidly settles back into a relatively planar disk. Gas
clumpiness can reduce the ISM cross section for self-interaction and thereby increase the dissipation timescale, but the supersonic turbulence that accompanies this clumpiness is highly dissipative. For modest clump filling factors, the net result is a greater dissipation rate for ISM kinetic energy. This is unlike the situation in stabilized ISM models, where the gas is supported by a steady thermal
pressure. Excess thermal energy can be radiated away, but the EoS, high
temperature floor, and/or feedback recipe, keep the gas warm in these
models and make the thermal support non-dissipative in practice.
Furthermore, the massive clumps that form in highly turbulent gas rapidly
migrate inwards through dynamical friction, unlike the relatively
homogeneous gas in stabilized models which settles into a circular orbit.

\medskip

\subsection{Interpretation: increased turbulence and rapid dissipation in mergers}

{\bf The effect of turbulence dissipation on the size evolution of a 
merger can also be estimated through simple calculations. A star and 
gas system in equilibrium satisfies the Virial Theorem, and then the 
total energy from self-gravity and motion equals half of the 
gravitational potential energy, $E_{\rm tot}=0.5E_{\rm grav}$, both of 
which are negative. Loss of energy by dissipation leads to a more 
negative $E_{\rm tot}$, and so a smaller, more tightly-bound 
equilibrium.  For typical mass distributions in a galaxy, the 
gravitational potential energy scales with the inverse of the radius, $R$, 
in which case the relative contraction is $\Delta R/R=-\Delta 
E/E=\Delta E/|E|$. 

If the gas mass is $M_{\rm gas}$ and the 1D gas velocity dispersion is 
$\sigma$, then $\Delta E=-1.5M_{\rm gas} \sigma^2$ when a 
significant fraction of the gas kinetic energy is lost. With similar 
notation for the stars, the total energy is $E_{\rm tot} =1.5M_{\rm 
gas} \sigma^2 + 1.5M_{\rm stars} V^2$, assuming that stellar motions are of the order of the circular velocity $V$ (even if not consisting in organized rotation). 
Thus $\Delta E/E=-X/(1+X)$ where $X=\left(M_{\rm gas}/M_{\rm stars}\right) \times \tau^2$, with $\tau = \sigma / V$. This ratio, 
$-X/(1+X)$, is also the relative contraction of the radius after all of 
the initial gas motion dissipates.  If $f_g$ is the disk gas fraction, 
then $X=\left(f_g/\left[1-f_g\right]\right)\tau^2$. 

Turbulent dissipation should occur in about a crossing time for radial 
motions \citep{maclow99}. The crossing time for a whole galaxy is 
about the orbit time, typically 100-200~Myr, so most of the 
non-rotational gas kinetic energy can be dissipated over the duration 
of a major merger. After dissipation and galaxy contraction, the gas 
receives new kinetic energy from the potential energy, and can then 
dissipate even more energy.  This cycle of dissipation, contraction, 
heating and more dissipation, continues with ever decreasing orbit time 
until the radial component of the turbulent energy is gone. 

Low-redshift Milky Way-like disks have $\tau \sim 3-5\%$, and their 
major mergers have a value about 3-5 times higher, $\tau \sim 20\%$ 
(observations: \citealt{irwin94,E95} -- simulations: 
\citealt{bournaud08,teyssier10}). Together with $f_g$ of a few percent, 
the size of a merger remnant is not much smaller than the original 
galaxy size. The same would be true in high-redshift merger models that 
have subsonic gas, i.e., gas without much energy dissipation, even if 
$f_g$ is large. 

High-redshift disk galaxies have $\tau \sim 20\%$ (e.g. \citealt{FS09}), $f_g\sim0.5$ (e.g., Daddi et al. 2010, Tacconi et al. 2010), and 
highly dissipative gas. With an increase of gas dispersions of a factor three during interactions (conservatively based on the known increased at low-redshift, but also consistent with high-redshift mergers spectroscopy in, e.g., Shapiro et al. 2008), $\tau$ would become of the order of $\sim 60\%$ for high-redshift mergers. Then $\Delta R/R\sim 0.25$. This result implies that turbulent dissipation for a gas-rich system can cause significant radial contraction of both the gaseous and stellar components for every gas dissipation time, i.e. just 100-200~Myr. As the gas continues to dissipate energy, the systemic contraction of the galaxy continues, and the total contraction of the merger can be larger if some mass is expulsed from the system and carries more energy away.

\medskip

Another mechanism for galaxy contraction is inward clump migration through dynamical friction. The timescale for this, given the clump masses and total masses of our models, is typically around 500~Myr (see Bournaud et al. 2007). This is slower than gaseous turbulence dissipation, and therefore not as important for overall shrinkage when the gas fraction is large. 

\medskip

Contraction also occurs if we consider the size of a disk to be determined by angular momentum.  Baryons in the central few kpcs of interacting galaxies loose significant angular momentum, while baryons initially in the outer disks carry away a large fraction of the initial angular momentum in tidal tails \citep[see review in][]{bournaud10m}. At first order, the angular momentum of baryons near the corotation region, which is typically about the half mass radius (a few kpc for the disk masses considered here) is be roughly conserved. 

For a mass $M$ encompassed by a radius $R$, and a circular velocity $V=\sqrt{GM/R}$, angular momentum conservation implies that: 
$$R \propto 1/VM \propto M^{-3}.$$ 

If gas initially outside the half-mass radius dissipates its turbulent 
energy and moves towards the center, the central mass increases, 
possibly as much as a factor of 2. Then the half-mass radius can 
decrease by a factor of 8, all the while conserving angular momentum. 
Further shrinkage can be caused by angular momentum removal through 
tidally expelled baryons. Only this tidal removal is present in models 
with a smooth, non-turbulent ISM. }

\subsection{Star formation history}\label{sec:sfr}

\begin{figure*}
\centering
\includegraphics[width=5.5cm]{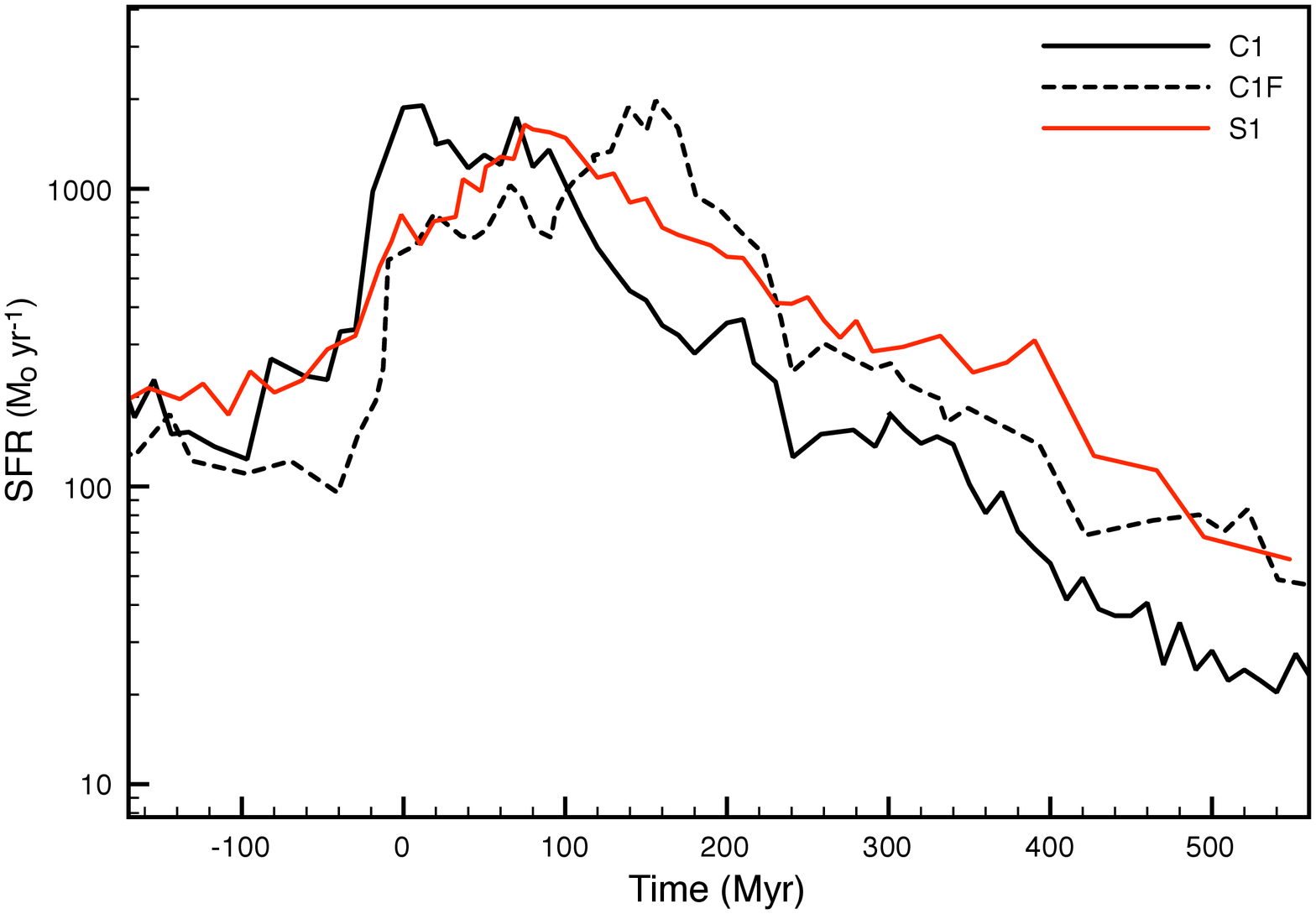}
\includegraphics[width=5.5cm]{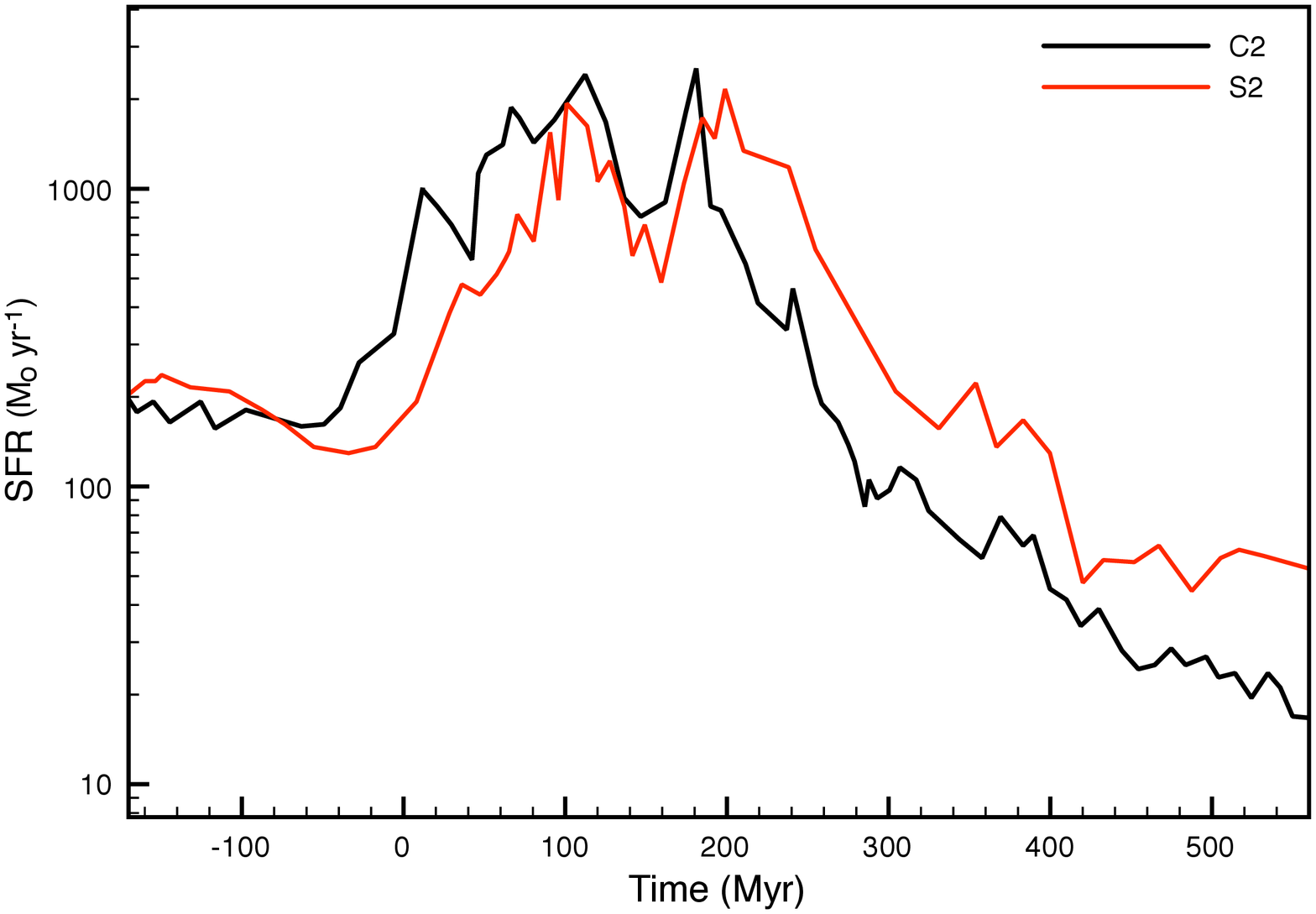}
\includegraphics[width=5.5cm]{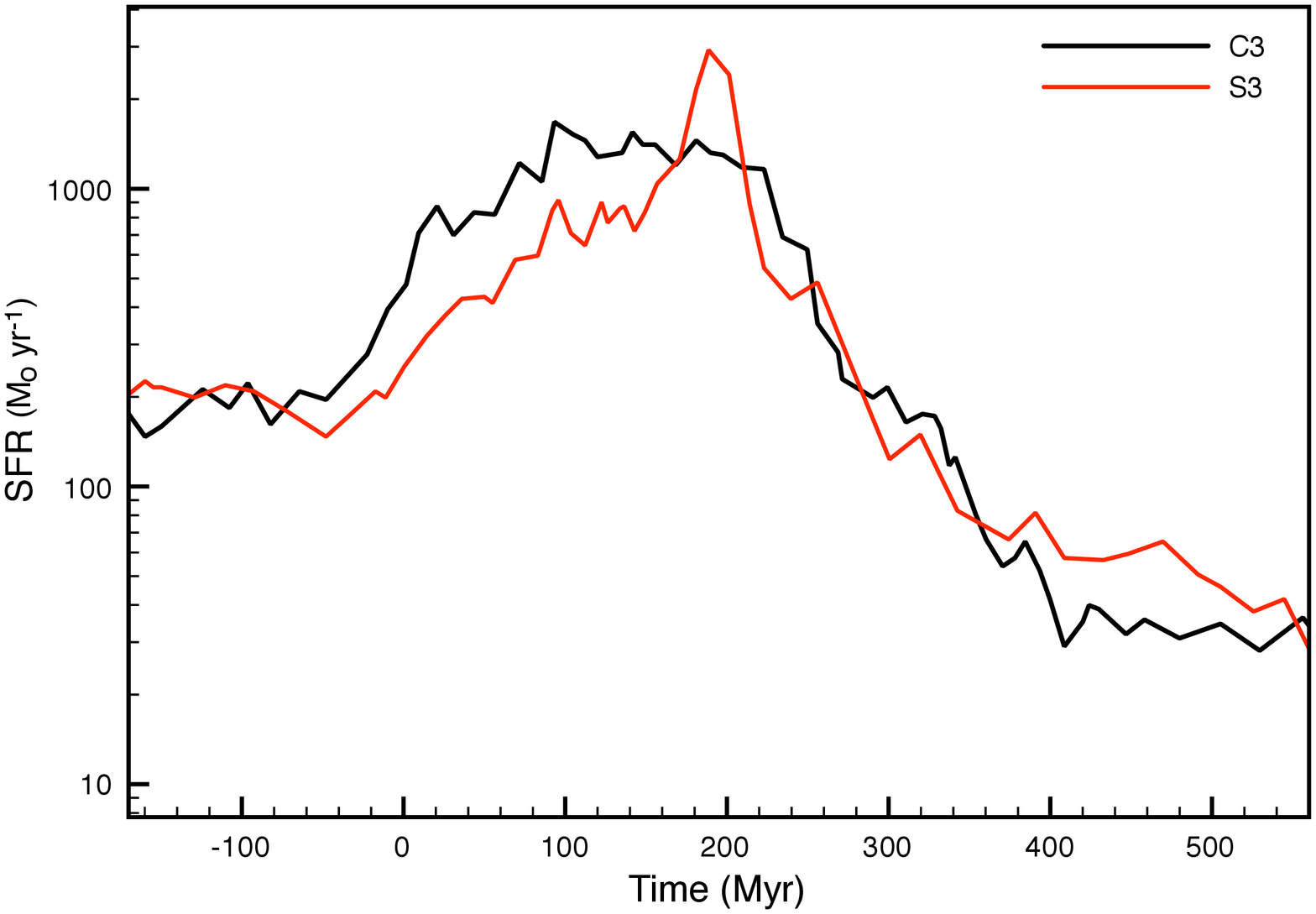}
\caption{Star formation history for the merger models along orbits 1, 2
and 3, with various EoS and feedback parameter. The pericenter passage
corresponds to $t=0$~Myr in each case.} \label{fig:sfr}
\end{figure*}

The star formation histories of mergers C1, C1F and S1 are shown in
Figure~\ref{fig:sfr}. The efficiency was calibrated in each model to
result in a SFR of approximately 100~M$_{\sun}$~yr$^{-1}$ in each
pre-merger galaxy, typical for $z \sim 2$ disk galaxies with similar
masses. The star formation rate during the merger reaches a factor of
$\sim$10 higher than in the pre-merger pair in all cases, with peak
values of 1000-2000~M$_{\sun}$~yr$^{-1}$. This rate is consistent with the SFRs
estimated in SubMillimeter Galaxies \citep[SMGs;
e.g.,][]{tacconi08}. The star formation peak can occur earlier-on or later-on
in various models, but the maximal SFR and the duration of the starburst
remain comparable, leading to globally similar amounts of
gas-to-star conversion for the cooling and stabilized models. The fraction of remaining
gas at the first pericenter and 300~Myr later are given in Table~2: no
major or systematical differences are found between the cooling and the
stabilized ISM models for any given orbit.

Both the cooling and the stabilized ISM models reach rates of star
formation that are realistic for observed $z \sim 2$ disks and
starbursting mergers. The rates are also consistent with
each other. This does not mean that gas cooling below $10^4$~K in a
clumpy turbulent ISM has no effect on star formation in
mergers. High-resolution models explicitly including gas cooling
could, in principle, resolve the densest gas phases that are actually
star-forming \citep{tasker, agertz, bournaud10}, but models using an
artificially thermalized ISM do not reach such high gas densities and they
have to describe star formation as a low-efficiency process pervasively distributed throughout a low-density gas. The calibration of the efficiency is subjective when dense clouds are not made self-consistently by the model. Here we choose an efficiency in the stabilized ISM model that gives the same star formation rate as the cooling model for an isolated disk. Had we used the same prescription for star formation in each case, then the cooling models would have had much higher star formation rates and their collisions would have been more gas-free.
{\bf While we compare the cooling and stabilized models with similar SFRs and resulting gas fractions, the EoS used to model cold gas imposes most of this gas to be relatively cold $< 10^4$~K. Different feedback models might result in higher amounts of hot extended gas, which could potentially cool down later-on and accrete onto a re-formed disk (see discusion in Sections~3.4 and 3.5.}

{\bf Globally, the star formation rate in our merger models peaks at about $\sim 10$ times the value of the pre-merger disk pair (with both the cooling and stabilized EoS). This factor is somewhat higher than the typical value in larger samples of merger simulations \citep[e.g.][]{dimatteo07,dimatteo08,MB08, stewart08} but not in large disagreement. This factor of about 10 also appears somewhat higher than in observations of mergers and close pairs \citep{bervall03, bell05, jogee09, robaina10}. However, these observations indicate the SFR enhancement at a random instant of an interaction, and what they consider to be major mergers are not strictly equal-mass ones. The factor 10 at the peak SFR in our models, which is reached only for a short period, is broadly consistent with the factor of 3-4 indicated by these observational studies. We did not here focus on particularly efficient cases where the SFR enhancement in a merger can reach factors larger than a hundred (as in \citealt{MH96} or \citealt{springel}), but rather on random, representative orbits.}

\medskip

The peak in the merger-induced star formation rate
begins somewhat earlier in the cooling models than in the
stabilized ones, just after the first pericenter passage
(Fig.~\ref{fig:sfr}). This earlier gas consumption cannot cause the
lower disk fraction of the merger remnants because {\em (i)} it occurs
only once the interaction process starts, so the gas fractions during
the interaction remain quite similar in both models (Table~2) and {\em
(ii)} the increased feedback in model C1F delays the peak of star
formation until a bit later than in model S1, but the final merger
remnant is almost as compact and disk-free as for the initial model C1.

The earlier merger-induced star formation activity in the cooling
models is consistent with the results obtained by \citet{teyssier10} in
low-redshift merger models \citep[see also][ for resolved star-forming regions in early pahses of galaxy interactions]{saitoh}. High-resolution models with a low
temperature floor can capture ISM turbulence, local dense shocks, and fragmentation into star-forming clouds. Stars form in these clouds when they are still in the main disk, long before they have accreted to the galaxy center. Such modelling is required to reproduce correctly the extent of star formation in early- and late-stage interactions, which is missed by standard prescriptions \citep[see also][]{barnes04,chien10}. On the other hand, low-resolution or stabilized ISM models get intense star formation only in the dense centralized gas that follows the merger-driven inflow. Dense clouds do not form in the disk. The delay in star formation for this stabilized case is comparable to the rotation period of the entire galaxies.

\subsection{Feedback and clump evolution in mergers}

Our simulations include a kinetic model for the energy feedback from
Type~II supernovae. Other sorts of feedback such as Type~Ia supernovae, stellar winds, and mass return are important for high-redshift galaxy evolution \citep{agertz10,MB10}, but the timescales for this feedback are typically longer than the timescale for a major merger of massive disk galaxies. Another sort of feedback with a short timescale is radiation pressure from young massive stars. Giant clumps in high-redshift, gas-rich galaxies could be sufficiently massive to self-regulate their star formation so that they ``survive'' feedback for several hundred Myr \citep{KD10}.This however remains debated \citep{murray}. The ``survival'' definition of \citet{KD10} implies that a bound stellar clump remains but not necessarily that the gas is retained by this clump. Expelled gas could be re-accreted by the clump and start forming stars again \citep{pflamm}.

These additional feedback mechanisms are not present in our models. Nevertheless, we found that varying the efficiency of the supernovae feedback (in runs C1F and LM-C2F) did not result in major changes to the properties of the final ETGs. Models C1 and C1F also have similar line-of-sight velocity dispersions in their gas component (Fig.~\ref{fig2} and \ref{fig3}). The main source of turbulence in high-redshift disk
galaxies is generally considered to be gravitational energy released
through clumping instabilities and/or inward mass accretion
(\citealt{EB10}, \citealt{DSC09}). {\bf Observations have suggested that the local velocity disperion scales with the local gas density and/or star formation surface density (\citealt{lehnert}, see also \citealt{green} for low-redshift analogues) which could suggest energy input from stellar feedback. The local velocity dispersion scales with the local surface density also in models dominated by gravity-driven turbulence \citealp[such as those in][]{BEM09}. Gas turbulence can be further increased by tidal forces in mergers \citep{E93,irwin94,E95,BDE08,teyssier10}. Stellar feedback is thus not expected to be the main energy source of the turbulent motions. The same appears to be true in local galaxies \citep{bournaud10}.}

While the pre-merger disk galaxies are already clumpy, the main gas clumps during the merger tend to be denser and more massive (see for instance the time sequence on Figure~1): some are pre-existing clumps that accrete more mass, some new clumps form with high masses allowed by the high gas densities and velocity dispersions. This maintains a clumpy stellar morphology during and after the merger. Most clumps are nevertheless not long-lived throughout the merger process. The average stellar age of the five largest stellar clumps in the final ETG in model C1 was measured 450~Myr after the pericenter passage, which is about 200~Myr after the final coalescence. These ages range from 80~Myr to 175~Myr. Hence these clumps are relatively young compared to the merger and formed in the latest phases. This suggests that our results do not directly rely on a low feedback efficiency, but primarily result from a strongly turbulent ISM, in which massive clumps naturally arise. 

{\bf We have not explored the modeling of other sources of stellar feedback (radiation pressure, winds, etc). We already discussed models with an efficiency up to 100\% in our supernovae feedback scheme, but Agertz et al. (2010) suggested that higher efficiencies, up to 500\%, could be more realistic -- presumably to account for other sorts of feedback that are not modeled, such as photoionization, or radiative feedback, the role of which at high redshift remains debated \citep[e.g.][]{murray, KD10, genel10}. Thus, we performed tests with SN feedback efficiency increased to 500\%, for the cooling and stabilized models on orbit~1. The results (see Table~2) show some variations, as stronger feedback fuels more extended and hotter gas reservoirs in the early phases of the merger, which adds mass to a re-formed disk component in the post-merger phases (following the process discussed, e.g., by Springel \& Hernquist 2005 and Governato et al. 2009). Nevertheless these variations are relatively minor compared to the initial discrepancies between models C1 and S1, both in terms of final disk fraction and final compactness. Furthermore, the differences between models C1F and S1F, or C1F5 and C1F5 respectively, are about of the same amplitude as the differences between the initial models C1 and S1. This suggests that, while a stronger feedback can moderately increase the disk fraction and half-mass radius of the final ETG, this effect is decoupled from the impact of using a cooling model generating turbulent gas rather than a thermally stabilized model. It does not strongly impact our previous conclusions on the effect of resolving gas cooling and strong ISM turbulence on the stellar size evolution and disk fraction in the final merger remnants.}

\subsection{Mergers of lower-mass galaxies}

Our main set of simulations used galaxy models that were relatively massive (without being unrealistic) for the typical disks and spheroids observed at $z \sim 2$. 

Lower-mass disk galaxies at $z \sim 2$, with circular velocities of 100-200~km~s$^{-1}$, also show giant clumps of star formation and gas turbulent speeds of several tens of km~s$^{-1}$ \citep{FS09, wright09}, like the pre-merger galaxies in our models. ISM turbulence and clumpiness should thus have qualitatively similar effects in high and low mass major mergers. This can be explicitly checked using models LM-C2, LM-S2 and LM-C2F.

\begin{figure}
\centering
\includegraphics[width=7cm]{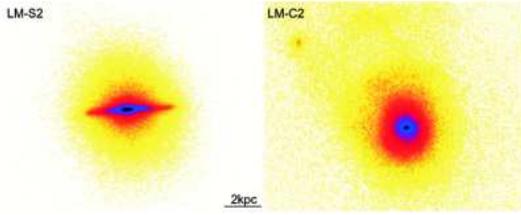}
\caption{Outcome of the lower-mass merger models LM-S2 and LM-C2 after full relaxation of the merger remnant. The projected stellar surface density is shown in each case, with an edge-on orientation with respect to the stellar spin axis. Although a weak disky distortion is observed in the cooling model LM-C2, only the stabilized ISM model LM-S2 has a massive and extended rotating disk component, and it also has a larger half-mass radius.} \label{fig:lowmass}
\end{figure}

The structural parameters of these low-mass mergers remnant are given in Table~2, and the edge-on stellar distributions shown in Figure~\ref{fig:lowmass}. The results are similar to what has been
detailed previously for mergers of higher-mass galaxies: the gas
velocity dispersions in the merging phase are higher in the cooling
model, and the merger remnant is significantly more compact, with only a low fraction of its baryons in a rotating disk, while a significant
disk component is found in the stabilized ISM model. In detail, the relaxed ETG after the merger is comparatively more compact, with a stellar half-mass radius reduced by a factor of 2.5 in the cooling models compared to the stabilized ISM model, versus 2--2.2 for the more massive cases.

\section{Comparison to observed galaxy populations}

\subsection{Kinematical signatures and dispersion-dominated systems}

The mergers modeled in realistic high-redshift conditions develop very high gas velocity dispersions, and could thus resemble some dispersion-dominated systems (DDS) observed in $z \sim 2$ surveys. To further probe whether such mergers would be easily identified in observations, three of us experienced in high-redshift spectroscopic data (TC, BE, KS) performed an eye classification of the velocity and dispersion fields from Figure~3 with gas density maps, and from Figure~4 without the density maps. The test was done without knowledge of the nature of the modeled systems. {\bf This test aims at understanding whether the kinematics of the simulated mergers would be recognized as typical for major mergers by observers familiar with the typical signatures expected for mergers. Additional instrumental/observational effects such as limited resolution, sensitivity, and noise, were not accounted for: hence this test relates only to the physical properties shown on Figures~3 and 4. Real observations could in addition miss small-scale or low-flux signatures and make the identification of mergers even more uncertain.}

Model~I is always recognized as an isolated disturbed disk, with some suggestions that the top-left clump could be a minor merger. The stabilized-ISM merger models are most often identified as mergers (two-thirds of the votes), generally with a high confidence level that these are binary major mergers. The cooling merger models were classified as DDS of unclear nature (about one-third of the votes), DDS of merger origin (one-third), or merger (one-third). More than half of the votes suggesting a merger or DDS merger also suggested a group or multiple merger rather than a binary major merger. This confirms that {\em (i) } on-going mergers with clumpy turbulent gas could be classified as DDS, and {\em (ii) } typical high-redshift mergers have substantially different kinematics from low-redshift mergers or models with artificially stabilized gas, which have smoother velocity gradients along extended tidal tails.

Many observed DDS have small sizes, so the high observed dispersions could sometimes result from blending at low resolution, but seem real in general \citep{law09}. There are nevertheless DDS that are relatively massive, extended, with chaotic velocity fields. Examples include Q1623-BX453, Q17000-BX710, and others in \citet{law09} and in \citet{FS09}. These could be major mergers of initially clumpy and turbulent galaxies, but their merger nature has not been clearly recognized yet because they lack the usual known signatures of mergers.

Disk galaxies with lower masses tend to be even more turbulent compared to the rotation speed \citep[lower $V/\sigma$, e.g.][]{FS09}. Reaching dispersion-dominated phases in mergers of such galaxies should be easier at lower mass. This might explain why observed DDS generally have small stellar masses, even though they are too numerous to all be major mergers. Overall, DDS are often suggested to result from violent gassy collapse \citep{law09}, which our models show can occur in high-redshift mergers.

\medskip

{\bf The local turbulent speed of gas in our cooling models is typically 150-200~km~s$^{-1}$ (Table~1), but the projected line-of-sight dispersions can be up to 500~km~s$^{-1}$ for central clumps and  up to 350~km~s$^{-1}$ for other clumps. These very high dispersions arise on short phases, shortly after the interaction pericenter and about at the peak of SFR. These dispersions may seem higher than what is found in high-redshift spectroscopic surveys. Note however that our models C1 to C3 involve galaxies that are more massive than most objects targeted in spectroscopic surveys, so they naturally have higher circular, and higher dispersion velocities at fixed $V/\sigma$. Furthermore, these high dispersions would be blended out by seeing-limited observations without Adaptive Optics. Substantial samples of Adaptive Optics spectroscopic data really exist only for disk galaxies \citep[e.g.,][]{genzel10b}. Other high-resolution data are unlikely to have targeted nearly equal-mass mergers at the peak of their high velocity dispersion phase. Whether or not the peak gas dispersions in our models are too high for real $z \sim 2$ objects is thus presently unknown; if this is the case, the effects of turbulence dissipation and size reduction in major mergers might be somewhat more modest.}

\subsection{Submillimeter galaxies as on-going mergers?}

SMGs are among the most actively star-forming objects at high redshift, and hence generally proposed to be the most actively starbursting phases of major mergers \citep{tacconi08}. An interesting test is then to compare the gas dynamics in our models to observations of SMGs. The properties of a representative $z=2.3$ SMG, SMMJ2135-0102, have been mapped spatially owing to gravitational magnification by \citet{swinbank}. While the
global properties are reminiscent of a gas-rich major merger, star
formation takes place in a few giant clumps with sizes of 100-1000~pc,
as in our merger models (Fig.~\ref{fig1a}). Interestingly,
\citet{swinbank} and \citet{danielson} noted that the molecular gas
spectra have a multi-component shape. The spectrum is more complex than the usual double-horn profile observed for rotating gas disks (including clumpy high-redshift disks, Daddi et al. 2010), and is best fitted by a combination of four main emitting components.
Multi-component spectra in SMGs that are not just double-horn profiles were also suspected by \citet{knudsen}.

\begin{figure}
\centering
\includegraphics[width=7cm]{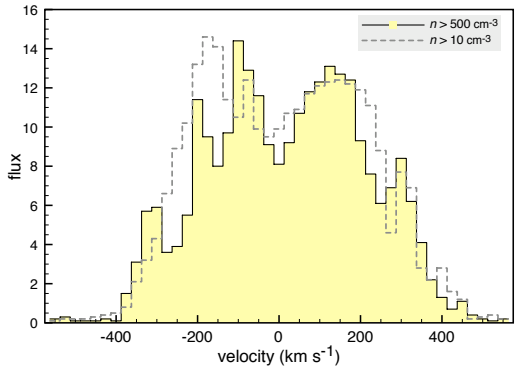}
\caption{Integrated spectrum of model C1, 55~Myr after the pericenter
passage, i.e., the second snapshot on Fig.~1, within a central aperture
of 4~kpc. Gas denser than 500~cm$^{-3}$ was selected as a proxy for
dense CO-emitting molecular gas. The dashed spectrum is for lower
density gas (denser than 10~cm$^{-3}$). The multi-component aspect of
the dense gas spectrum is reminiscent of the CO spectra observed in
SMGs (see text for details).} \label{fig:cospec}
\end{figure}

We modeled a ``molecular gas'' spectrum for model C1, 55~Myr after the pericenter passage, using gas denser than 500~cm$^{-3}$ as a proxy for molecular gas. The modeled spectrum, corresponding to the second snapshot on Figure~1, is shown in Figure~\ref{fig:cospec}. It has a multi-component aspect similar to that in the observation. Our models did not aim to reproduce this SMG in particular, and each individual component is not reproduced, but the amplitude of the emitting peaks is consistent with the spectra observed by \citet{danielson}. This property results from the dense cold gas (and star formation) being mostly located in a few giant clumps. Interestingly, the stellar mass at this instant in model C1 is $2 \times 10^{10}$~M$_{\sun}$, with a gas fraction of 40\%, close to the estimates for SMMJ2135-0102. The star formation rate in the model is 1400~M$_{\sun}$~yr$^{-1}$, somewhat higher than observational estimates for this object \citep{ivison} but not inconsistent with SMG-like activity in general. The modeled molecular gas spectrum has a full width at half maximum and a full width at zero intensity that are quite close to the observed values, suggesting quantitative agreement on global gas dynamics between our model and a typical SMG of similar mass. Furthermore \citet{danielson} noted that the denser molecular gas tracers (such as high-$J$ CO lines) have more peaked spectra, while lower-density tracers have smoother spectra
(Greve et al. 2005 also found relatively smooth low-$J$ spectra in
SMGs). This is consistent with the peaked components being associated
with dense star-forming clumps, and with our models showing smoother
spectra for lower-density gas (Fig.~\ref{fig:cospec}).
Higher-resolution simulations resolving cooling down to lower
temperatures would be required to qualitatively explore further this temperature
dependence of molecular gas dynamics in SMGs.

Standard models of high-redshift mergers using warmed stabilized gas to
model the ISM are only mildly successful in explaining the properties
of SMGs. They can model 805$\mu$m fluxes that account for SMGs
\citep{narayanan}, but only when the pre-merger disks are calibrated to
high submillimeter fluxes: the typical enhancement of the 850$\mu$m flux
during the merger/SMG phase is not larger than a factor of 2--3. These
models also have smooth gas kinematics. If dust follows the
scattering of gas in a three-dimensional {\bf clumpy distribution in and around
the centrally collapsed system, as in our models, then the obscuration of star
formation could be very high under some lines of sight}, like in SMGs \citep{thronson90}.

Another model of SMGs suggests that they are the high-mass end of
cold-flow galaxies, with no mergers involved \citep{dave10}. Some SMGs could be like this, particularly those with large radii, relatively low star-formation rates, and relatively long durations compared to
the orbit time. The SMGs in our model have much stronger bursts
and the disk gas is consumed more quickly than in the cold flow
models, so we expect a different proportion of elemental abundances in these two cases, and different star formation histories. Multi-component gas spectra could be a confirmation of a clumpy merger origin for SMGs, if they appear to be commonly observed in larger samples.

\subsection{Clumpy morphologies in high-redshift mergers and ETGs}

The stellar distribution in our models shortly after the pericenter
resembles a single ``clumpy galaxy'', but it is more asymmetric than
the pre-merger clumpy disk models  (see second and third panels in
Figure~1). This could be consistent with the population of ``assembly
galaxies'' in \citet{E07}, with entirely clumpy morphologies not
consistent with their being isolated disks. Other examples of
high-redshift clumpy mergers can be found in \citet{lotz06}.

Our modeled high-redshift clumpy mergers lack very prominent
tidal tails. There are long tidal features, but they are less
contrasted than the star-forming clumps. This differs from low-redshift
mergers, where long tidal tails with ample and smooth velocity
gradients are typical \citep[e.g.,][]{bournaud-fp}.

\medskip

Once the mergers are relaxed, stellar clumps are seen in the central
kpcs of the final ETGs (Fig.~\ref{fig1b}). They are relatively young
($1-2 \times 10^8$~yr in the final snapshots) and their masses are all
lower than $2 \times 10^8$~M$_{\sun}$ -- while our pre-merger disk
models harbor clumps of several $10^8$~M${\sun}$. A lower clump mass
is expected from stabilization of the merger remnant by the main
stellar spheroid \citep{martig09, DSC09}. Residual star formation is
associated with those clumps after the merger, consistent with
high-redshift red, but not completely dead, ETGs \citep{tonini}.

\begin{figure}
\centering
\includegraphics[width=7cm]{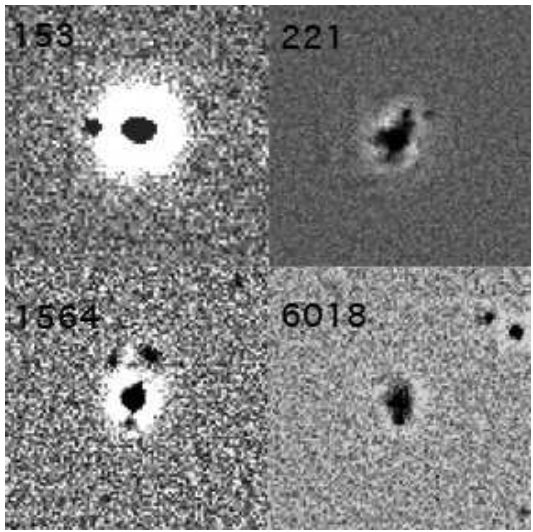}
\caption{Four examples of elliptical galaxies in the UDF where unsharp
masking unveils clumps in/around the main body of the ellipticals. Other
examples and details can be found in \citep{E05}. The clumps are
relatively young, and less massive than the clumps in UDF disk
galaxies, consistent with the clumps in early-type galaxies in the
outcome of our merger models. } \label{fig:UDF-clumps}
\end{figure}

These clumps might be harder to observe than the most massive ones in
$z \sim 2$ star-forming disk galaxies, but young stellar clumps have
actually been observed in the central body of high-redshift ellipticals
in the Hubble Ultra Deep Field \citep[UDF; ][ see examples in
Fig.~\ref{fig:UDF-clumps}]{E05}. Those clumps have young stellar ages,
and their masses are typically a fraction 10$^{-3}$--10$^{-4}$ of their
host galaxy masses -- lower than in clumpy disks but consistent with
our ETG models. Other clumps are also seen around several ETGs in the
Mancini et al. (2010) sample.

The outermost clumps, those formed at more than 50~kpc from the
center in model C1, could become dwarf galaxies formed during the
merger \citep[tidal dwarf galaxies (TDGs)][]{duc2000}. This would
lend support to recent suggestions that the rate of TDG formation could be high at high redshift \citep{kroupa-tdg}.

\subsection{Compact high-redshift ETGs}

Our merger models with cooling predict final stellar distributions that
are relatively compact. The properties of the final ETGs in our cooling
models are comparable with the properties of the three most compact high-redshift ETGs in
the \citet{mancini} sample with similar masses, as well as some objects
in \citet{cappellari09}. Faint extended stellar halos with high Sersic
indices are present in the observations, like the one surrounding the massive kpc-sized
core in the final ETG of model C1 (Fig.~4).

Real mergers should involve various mass fractions, gas fractions, and
degrees of stability and clumpiness in disk galaxies. Thus, they should form a variety of ETGs, ranging from very compact objects \citep[as spectroscopically confirmed,][]{vdk,cappellari09} to more extended ones (as present in the Mancini et al. sample). If high-redshift galaxies with lower masses are relatively more turbulent \citep{FS09}, then their mergers could be even more dissipative than the more massive disk galaxies. Then the most compact ETGs would be the lowest-mass (Section 3.5). Such a trend is observed among the various samples of high-redshift spheroids (see Fig.~4 in \citealt{mancini}).

\cite{wuyts} also proposed recently that high-redshift mergers lead to
the formation of compact ETGs. Their simulations have a
stabilized, pressurized ISM model (although implementation details
differ from our ``stabilized'' models). There is some compacting of the system during the merger, but at the same level as in our stabilized models and not at the level of our cooling models of clumpy galaxy mergers. Actually, to form ETGs that are about as compact as $z \sim 2$ ones, \citet{wuyts} had to start with pre-merger disk galaxies that were twice as compact as our initial disk galaxies, with a disk scale length half as large. {\bf Which initial conditions are more realistic in terms of disk sizes is unknown. The known size evolution between $z \sim 2$ and $z =0$ \citep{franx} mostly concerns spheroids. \citet{bouche07} propose that high-redshift {\em disks} have the same velocity-size relation as disk galaxies at $z=0$ and a spin parameter 2 to 6 times larger than in the Wuyts et al. initial conditions, but this was called into question by \citet{dutton10}. Our initial disk galaxies still seem realistic for massive star-forming disks at $z \sim 2$}: a radial scale-length of 5~kpc for a circular velocity of 245~km~s$^{-1}$, to be compared for instance to the sub-sample of disk galaxies in \citet[][Fig.18]{FS09}. 

It thus appears that a major merger model can produce the transformation of a typical $z \sim 2$ star-forming disk galaxies (hence relatively extended) into a typical high-redshift ETG, only if ISM cooling, turbulence and clumpiness are captured in the simulation. Observations do suggest that compact ETGs formed from relatively diffuse disks, rather than disk galaxies that are already compact \citep{ricciardelli}. It nevertheless seems that our models predict Sersic indices that are somewhat higher than those estimated for compact high-redshift ETGs. In observations, \citet{puech06} also noted that compact luminous galaxies at high redshift rarely show rotational support, but instead have large dispersions and chaotic velocity fields. They suggested that high-redshift mergers produce compact remnants.

\subsection{Long-term evolution towards low-redshift ellipticals}

High-redshift ETGs probably do not survive in the form of compact
objects in the nearby Universe. They should grow mostly through minor
mergers, which can increase their stellar sizes \citep{naab09}. In this
dry process, the small and compact disk components remaining after
high-redshift mergers (Fig.~1 at $t=625$) could undergo inefficient
star formation \citep{martig09} and persist as compact disks of
residual gas and stars, as often found in the central kpc of nearby
ellipticals \citep[][]{young08,crocker10}.

Of course some of these high-redshift ETGs could accrete external gas in sufficient amount to reform a bright and massive stellar disk, in a ``spiral rebuilding'' scenario, as proposed for instance by \citet{puech06}. However our models show that such disk rebuilding does not happen spontaneously from gas expelled during the merger (in tidal tails and/or through feedback processes) that would later on fall back onto a disk. Large accretion of external gas would be required to rebuild a massive star-forming disk galaxy. Models by \citet{martig09} have shown that this can naturally happen in some haloes in $\Lambda$-CDM cosmology, when the late mass assembly history is dominated by smooth infall rather than mergers.

\section{Conclusions}

The ISM is very turbulent and clumpy in high-redshift galaxies.
Simulations of high-redshift mergers with cooling and sufficient resolution
to develop turbulence and dense gas clouds show major differences
compared to models with smooth, thermally supported disks:

\begin{itemize}

\item On-going mergers of representative high-redshift galaxies can
have very irregular gaseous spheroids during the merger, with clumps
and filaments throughout, instead of rotating gas disks surrounded by
long tidal tails with ample and smooth velocity gradients, as is common in low-redshift mergers.

\item The gas velocity dispersion in these mergers becomes very high.
At some stage, most of the gas kinetic energy is in the form of a
three-dimensional velocity dispersion. Such mergers could take the
appearance of dispersion-dominated galaxies with high gas dispersions and chaotic velocity fields. Their spectral properties are also consistent with SMGs.

\item Turbulent and clumpy gas undergoes a violent and dissipative
collapse in mergers. The final ETGs in these mergers are relatively
compact, with high Sersic indices and faint outer stellar halos.

\item  The masses and sizes of disk components that survive or re-form after major mergers are strongly reduced when the hydrodynamics of the cold turbulent ISM is taken into account. Large disks at high redshift should form mainly through cold accretion rather than mergers.

\item {\bf Strong feedback processes can somewhat increase the size and disk fraction of the final ETGs formed in these mergers, but the impact of directly modeling a cold turbulent ISM phase rather than using a thermally pressurized model remains independent from the efficiency of stellar feedback.}

\end{itemize}


Wet mergers at high redshift should involve galaxies that have
turbulent and clumpy gas-rich disks. Mergers of such galaxies can
undergo dispersion-dominated phases. They can also produce starbursts with SMG-like properties. The late phases are accompanied by dissipative collapse and the formation of compact ETGs. This is consistent with the evolutionary sequence proposed, e.g., by \citet{ricciardelli}. Massive rotating disk components cannot easily survive such mergers, a property that is not reproduced correctly by simulations where ISM turbulence and clumpiness are unresolved and modelled with a thermally pressurized equation of state. Large rotating gas-rich disks at high redshift, such as those observed by \cite{genzel06} and others, are probably not the outcome of a
mass assembly dominated by mergers. {\bf Large far-outer reservoirs of gas, pre-existing to mergers or fed by strong feedback during the mergers, could participate to some disk rebuilding (as observed in the \citealt{governato09} model). Stellar mass-loss and gravitational torquing over cosmological timescales can also help preserve large disks (as in the \citealt{MB10} model, see also \citealt{leitner}), but continuous cosmological infall of fresh gas probably dominates the assembly of disk-dominated galaxies down to $z=0$.}

If gas-rich mergers at high-redshift are characterized by strong turbulence, the spread of the gas density distribution could be very large, as it scales with the turbulent mach number \citep{KT07}. Not only the average gas density would be high, but there would also be an excess of dense gas at number densities of thousands per cm$^3$ and more, compared to non-interacting systems. Dense gas excess has already been found from low-redshift merger models \citep[e.g.,][]{juneau09,teyssier10} and could explain the enhanced HCN/CO ratios in nearby ULIRGs. An excessive fraction of dense gas could also increase the star formation rate at fixed average gas density, possibly consistent with recent observations of star formation in disks and mergers suggesting enhanced efficiency of gas consumption in mergers \citep{daddi10b, genzel10}.

\acknowledgments We used HPC resources of CINES under GENCI allocation 2010-GEN2192. We acknowledge useful comments from Emmanuele Daddi, Eric Emsellem, Reinhard Genzel, Philip Hopkins, David Law, Paola Di~Matteo and the referee, and are grateful to Ian Smail and Mark Swinbank for suggesting comparisons with SMG spectral properties. This work was supported by the Agence Nationale de la Recherche under contract ANR-08-BLAN-0274-01.

{}

\end{document}